\documentclass{article}

\usepackage{geometry}
\usepackage{doc}

\usepackage{url}
\usepackage{wrapfig}
\usepackage{graphicx}
\usepackage{epstopdf}
\graphicspath{ {images/} }
\title{A Cost-based Placement Algorithm for Multiple Virtual Security Appliances in Cloud using SDN: MO-UFLP(Multi-Ordered Uncapacitated Facility Location Problem)}

\author{
	Dana Jamaluddine \thanks{Concordia University,Montréal, QC H4B 1R6,Canada , danajamaleddine@gmail.com}
	\and
	Shalaka Kolhapure\thanks{Concordia University,Montréal, QC H4B 1R6,Canada , shalaka.kolhapure@gmail.com}
	\and
	Prajeesh Murukan\thanks{Concordia University,Montréal, QC H4B 1R6,Canada , p\_muruka@encs.concordia.ca}
	\and
	Fady Mikhael\thanks{Concordia University,Montréal, QC H4B 1R6,Canada , f\_mikh@encs.concordia.ca}
	\and
	Shiva Nouzari\thanks{Concordia University,Montréal, QC H4B 1R6,Canada , s\_nouza@encs.concordia.ca}
}
\date{}

\begin{document}

\maketitle

\abstract{
Software Defined Networking (SDN), has introduced many advanced platforms for managing networks and adopting different security tools with them, but the cost of these platforms should be considered as well.
In this paper, we present an extension of the existing approach to the optimal placement of virtual security appliances in a pre-defined network setting. The approach proposed by Bouet [1] only considered one security appliance, we extended his approach to several virtual security appliances. We conducted several simulation tests showing good performances of our approach. To show the feasibility, we implemented our approach using SDN and virtual security appliances and integrated it into OpenStack. This extension adapts UFLP algorithm to real world situations where several middle boxes need to be deployed to satisfy security needs for the applications deployed in the cloud.
We realized this approach by implementing "OpenStack on top of OpenStack" , a nested OpenStack implementation with OpenDayLight as the SDN controller . 

}

\pagebreak
\tableofcontents

\pagebreak
\listoffigures

\pagebreak
\listoftables

\pagebreak

%
%
\section{Introduction}
\label{introduction}

In today’s IT systems, cyber security requires fine-grained, flexible, adaptable and cost optimized monitoring mechanisms. The emergence of new networking technologies, such as Network Function Virtualization (NFV) and Software Defined Networking (SDN), opens up new venues for large scale adoption of these cyber security tools. In particular, Security Appliances can be virtualized and dynamically deployed as pieces of software on commodity hardware. Deploying such software Security Appliances is costly in terms of license fees and power consumption. Designing cost effective Security Appliance deployment strategies that meet the cyber security operational constraints is thus mandatory for the adoption of this approach. For this purpose we propose a method, based on genetic algorithms, that optimizes the cost of Virtual Security Appliance deployment, minimizing their number, the global network load and the number of unanalyzed flows. We conduct several experiments with different types of traffic and different cost structures. The results show that our approach is able to reach a trade-off between the number of virtual security appliances and network load.

\subsection{Problem Statement }
One of the interesting work on cost-optimization security deployment is the work of Bouet [1]. The latter focus on the optimal placement problem of Deep Packet Inspection (DPI) engines based on the Uncapacitated Facility Location Problem (UFLP). To tackle the scalability problem, Bouet proposed the use of genetic algorithms.Bouet defines[1] the problem in such a way that for a given network infrastructure and a given traffic matrix, find a Security Appliance deployment that minimizes the overall cost of the deployment. This cost is the result of a joint optimization that minimizes i) the number of Security Appliances, ii) the overall network load induced by flow redirections through the Security Appliances, and iii) different operational constraints. These constraints concern financial costs such as the cost associated to a deployed Security Appliance (e.g. license price, CPU utilization, energy consumption...), the cost associated of network resources (e.g. network total cost of ownership, capacity of the network to absorb new traffic), and the cost of penalties due to the incapacity to analyze a flow. The constraints also include management limits such as maximum number of engines to be deployed, the maximum used bandwidth per link (to be able to absorb peaks) and the maximum unallocated flows. The two main objectives, that are minimizing the number of Security Appliances and minimizing the network load, are in fact orthogonal. Indeed, all the flows have to go through at least one Security Appliance to be analyzed. When the number of Security Appliances is small, the paths tend to be elongated. Therefore, minimizing the number of engines increases the additional used bandwidth. On the contrary, minimizing the used bandwidth increases the number of Security Appliances to be deployed.
\subsection{MO-UFLP Contributions }
Bouet has focused on two important goals[1]. The first one is to reduce the number of security machines, and the second to make sure all the traffic goes through the selected security machine. His approach has clearly met these goals but there are some points which have not been taken into consideration.
Bouet's algorithm has been designed in such a way that the output will give you only one place for flow.
Also as there is only one security machine, the order of the deployment among several security appliances has not been considered .
To address this problem, MO-UFLP aims at considering several instances of different types of virtual security appliances.  

To this end, We propose to extend Bouet's approach to take into account the number of security machines which are intended to control the traffic and decides on the optimal placement and optimal number of these security appliances.The other major added value  of our approach was to consider the order of the security machines in the algorithm. In section 2.1 we will detail partly Bouet approach together with our extensions. In section 2.2 we describe our algorithm and how it works . Section 4 shortly compares different type of implementation on different topologies. Section 5.1 defines the OpenStack services that was used, the next part describes how the environment was installed and prepared. The rest of section 5 focuses on how the scheduler and genetic algorithm has been implemented and were connected to each other and also describes about their input and output files. Then its followed by related works which talks about experiments which are related to this paper. Conclusion and references are the last parts of this paper.

\section{MO-UFLP Approach}
\subsection{Extension to Bouet Work}

 A cloud network is considered as a graph, where each node is an area that could potentially hold security machines. We want to deploy different types of security machines in any node. Each of these security machines has a specific cost. For the time being, we consider all the types to have the same cost. We  take into consideration that from any source we can find a flow to any destination that goes through all the security machines types at least once and in an ordered fashion. The penalty of unanalyzed flow is given to the cost due to an inability to find such flow.
 We can also add a threshold for the maximum number of security machines that can be deployed, but for now we discard this option because we are currently not interested in limiting the number of security machines. These criteria, along with the length of the path, form a fitness value. A lower fitness value represents a better solution since we are trying to minimize the cost.\\ \newline

 Thus we formalized the problem  described as follows:\\
 •n is number of nodes in the topology, x is an array of n numbers [0, 2, 1, 3, 0, 1...], which represents the presence, or absence of security machine in a particular node. 0 represents the absence of a security machine, where any other number will represent one type of security machines. The global cost function is represented by the Fitness Function F (x):\\ \newline
$ F(x) = fSM (x) + fpath(x) + funalloc(x) \\
 fSM (x):$ Represents the cost of a security machine deployed in the network.\\
 $fSM (x) = SMcost * n(x) $\\ \newline
 
 n(x) is the total number of security machines used in the graph. We can define a threshold value for the total number of security machine that indicates the maximum number acceptable to be deployed in the network. If the solution contains more than that threshold value, we will discard the solution and go to next possible solution. For the time being this number is set to the maximum.\\
 fpath(x): Total duration of the path\\
 Each edge between two nodes has a weight. This weight can either represent a bandwidth value or in our case a duration. This function adds all the time durations of each link included in the path. We will try to select the shortest distance path between source and destination, which goes through the security machines.\\
 funalloc(x): The sum of penalties\\
 funalloc(x) = penalty * u(x)\\
 u(x) is total number of unanalyzed flows. We will check if we can find at least one flow in the network from every source to every destination that will go through all the security machine in a specified order. If such a flow is not found, we consider the path between this sources to this destination as unallocated. 

\subsection{Multiple Ordered UFLP (MO-UFLP) Algorithm}
Our solution is based on genetic algorithm. Genetic algorithm is a search procedure based on evolutionary algorithms principles. Each solution in GA has a set of chromosomes, or genotype that will go through mutation. Individuals form a population that has generations depending on the number of iterations it goes through. In every generation, the fitness of an individual is calculated and the fittest individuals are selected to be the parents of the next generation. The algorithm terminates either when a specified number of generations are reached, or when it reaches an ideal fitness value for an individual.
Genetic Algorithm usually includes 4 steps:
\begin{itemize}
	\item  Initial population: The initial population in GA is randomly created. Each individual of this population is a potential solution for the problem. A population is called a chromosome and its elements (individuals) are the genes. A number represents genes and the chromosome is a set of gene representation.
	\item  Selection: At each iteration (generation) a group of solutions are selected to be the parents of the future Generation. The selection is based on the fitness value of the individuals in the solution. The computation of the fitness function and the number of parents selected will depend on the problem in hand.
	\item  Crossover: The selected parents will produce offspring that will inherent some of the parents’ genes. The crossover will swap information between a pair of parents. The simple crossover will swap the k first element from one parent with the k elements of the second parent. Two new chromosomes are obtained. The k depends on a crossover probability specified in the algorithm.
	\item Mutation:	Population is chosen and a new generation is reproduced.
	Finally, the mutation will alter some genes of the next generation based on a mutation probability. The resulting solution differs from the initial and it usually has a better fitness average since only the best (fittest) are chosen from the previous population. The section-crossover-mutation steps are repeated until a termination condition is reached.
\end{itemize}

 Any node can potentially hold a security machine. The algorithm will randomly allocate security machines in the graph by giving the genes values that range between 0 and the number of security machine type. The fitness value for each individual (potential solution) will be calculated based on the previous equations. Because the fitness value presents the number of security machines and a penalty for unallocated flow, the higher the value, the less optimal it is. \\
 Based on the following conditions:
 
 \begin{itemize}
 	\item The number of unanalyzed flows is less than threshold value.
 	\item the security machines are ordered in each flow.
 \end{itemize}

We will consider that solution, otherwise we will discard the solution and go to next chromosome. The two best chromosomes will constitute the parents of the next evolution.\\ \newline

\section{Numerical Results}
\subsection{Random Graph Experiments}
The implementation was tested on random graphs of size 10, 20 and 50 nodes[Table 1]. The graphs were tested for up to 5 security machine types, assuming we will not be in a situation that needs more than 5 security machines. The results are as follows: 
For each graph of 10, 20 and 50 nodes we tested for 1, 2, 3, 4 and 5 security machines types and we measured the time it takes for the genetic algorithm to complete. In all the tests 5 evolutions were tested. 

For example, for a graph of 10 nodes with 5 security machines types, we could not get any solutions after 5 evolutions. We increased the number of evolutions to 10. For the graph size 20 and 5 security machines, we could not find any solutions because the graph does not contain a path from source to destinations with more than 4 nodes in between.\\

\begin{table}
	\centering
	\begin{tabular}{|c|c|c|c|c|c|}\hline
		Node/SM & 1 & 2 & 3 & 4 & 5 \\\hline\hline
		10 & 89 & 99 & 133 & 145 & 255 \\
		20 & 105 & 173 & 303 & 316 & N/A \\
		50 & 183 & 262 & 215 & N/A & N/A \\\hline
	\end{tabular}
	
	\caption{Graph Comparison (Time in ms)}
	\label{table-sample}
\end{table}

\subsection{Fat Tree Topology }
The next evaluation was  realized based on a fat tree with k=4 [Figure 1]. We gave all the edges the same weight because we wanted the solution to find the shortest path that goes through the least number of security machines. Below Figure 1 represents the graph. In the fat tree, it was considered the first 8 hosts to be the source (0 to 7) and the last 8 hosts as the destination (8 to 15). For 1 type of security machine, the best solution is adding an SM in node 35. It can be seen that the genetic algorithm is always choosing the shortest paths from all sources to destinations. The results are shown in the table 2.\\
\begin{figure}[h]
	\caption{Fat Tree of Nodes}
	\centering
	\includegraphics[width=0.9\textwidth]{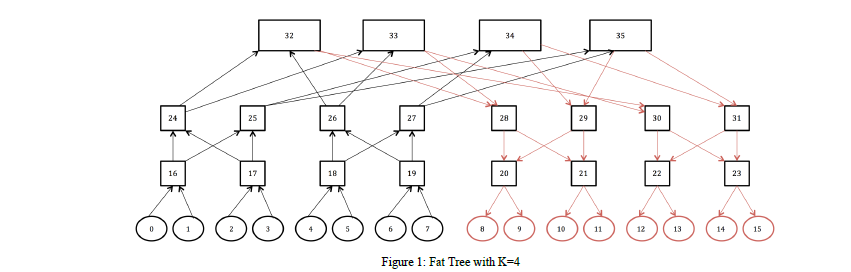}
\end{figure}
Since the graph has to be unidirectional for the time being, we ran the code on the fat tree twice; once in each direction.
\begin{table}
	\centering
	\begin{tabular}{|c|c|}\hline
		SM Type & 5 Evolution  \\\hline\hline
		1 & 555  \\
		2 & 502  \\
		3 & 547  \\
		4 & 803  \\
		5 & 771  \\\hline
	\end{tabular}
	
	\caption{Evolution Comparison in ms}
	\label{Evolution Comparison}
\end{table}

\subsection{Results and Discussion}

After many trials and changes, a point was reached where an optimized solution can be achieved in a short period of time. It was noticed that each time either the number of nodes or the number of security machine types were increased, it was harder to find a solution. This is because the genetic algorithm gives the genes random value and depending on the number of evolutions that we give, these random values either can pass all checks or the solution will fail. By increasing the number of evolutions or increasing the number of population, we are giving the algorithm more chance and time to find the best solution.\\

\section{MO-UFLP Implementation and Integration in OpenStack}

OpenStack is an open source software for building and managing public and private cloud computing. OpenStack let users deploy virtual machines and other instances that handle different tasks for managing a cloud environment on the fly. But most importantly, OpenStack is an open source software, which means that anyone who chooses to can access the source code, make any changes or modifications they require, and freely share these changes back out to the community at large. OpenStack falls in the category of Infrastructure as a service.[2]
The following services are used in the implementation phase:[3].\\
Nova is the primary computing engine behind OpenStack. It is used for deploying and managing large numbers of virtual machines and other instances to handle computing tasks. It supports variety of virtualization technologies like KVM, Xen and VMware. In Nova the queue, RabbitMQ by default is the central node for message passing.\\
Neutron provides the networking capability for OpenStack. Neutron server accepts the API- requests and maps them to correct plugins. It helps to ensure that each of the components of an OpenStack deployment can communicate with one another quickly and efficiently[4].\\
Horizon is the dashboard behind OpenStack.  Developers can access all of the components of OpenStack individually through a Restful application programming interface (API), but the dashboard provides system administrators a look at what is going on in the cloud, and to manage it as needed.

\subsection{OpenStack on Top Of OpenStack}

Realization of MO-UFLP needs a complex topology for verification and validation. For this purpose a 12 node OpenStack environment has been created using “OpenStack on OpenStack” methodology. In this methodology, a set of high performance hardware platform is chosen, and OpenStack virtualization techniques are used to create additional OpenStack controller and nodes. This platform is highly scalable and it gives a unique advantage to the implementation team to create diversified network topology. Given the hardware platform, the project team decided to create 2 layers of open stack installation over a set of 4 high performance blade servers , so that there will be adequate number of nodes and controllers. These nodes and controllers will be configured using SDN to perform configurable Software Defined network topology. 
To make the installation and topology simpler, a layered approach of open stack installation has been designed and implemented.  Layers are numbered as Layer-0 and Layer -1, in which the Layer -0 Openstack represents the initial installation of Open stack on top of the hardware directly (bare metal) or on top of an existing operating system. In our case, we decided to use a hosted open stack installation. Layer -0 Openstack installation is performed over 4 high performance blade servers, with Ubuntu 14.04 TLS server operating system installed.

\begin{figure}[h]
	\caption{Virtual Machines Layer-0}
	\centering
	\includegraphics[width=0.9\textwidth]{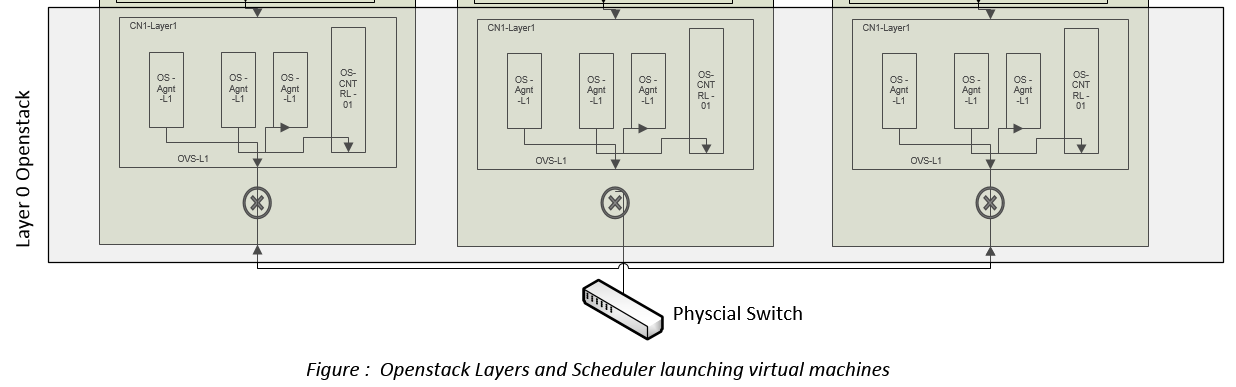}
\end{figure}
\subsubsection{Layer 0 OpenStack} 
All 4 blade servers were chosen to install hosted Openstack platform over Ubuntu 14.04 TLS. Blade-01 was chosen to install OpenStack controller and the remaining 3 blade servers were configured as open stack nodes. Since it is a highly customized environment, Openstack installation was done on all 4 blades using devstack. 
Openstack controller installation on Blade-01: blade 01 is chosen as controller, and it contains additional configurations to install Openstack dashboard, Nova network services, and other required Openstack components. 
Blade 02, 03 and 04 are configured as nodes. They have only minimal components such as nova agent and  neutron agent. Also, these nodes are connected to the Openstack controller via additional configuration options at the nodes. 
Blade-01, the server with open stack controller, is stacked first, followed by the nodes. Once all nodes and controller are stacked, the controller identifies the nodes and utilizes resources such as storage, memory and computing power of nodes. Once the open stack platform is up and running, additional virtual machines can be launched on the layer 0 platform.
We decided to launch 13 virtual machines on top of Layer 0 open stack platform with minimal computing and memory resources so that they can be installed with Openstack. In summary, Layer-0 Openstack runs on top of Blade servers (with Ubuntu 14.04 TLS), and VM are launched over this Layer -0 platform.\\
\subsubsection{Layer 1 OpenStack }
This layer is designed over the VMs launched on Openstack Layer-0. There are 13 VMs launched at Layer 0. One of them will be treated as controller for Layer 01 and remaining 12 will be designated as nodes.

All 13 VM are installed with Ubuntu 14.04 TLS. One of the VM is designated as a Layer 1 Openstack controller. This VM will also have OpenDaylight – SDN controller. OpenDaylight is integrated with OpenStack and this controller is responsible for routing and re-routing traffic between the VMs launched over Layer 1 Openstack. \\
\begin{figure}[h]
	\caption{Virtual MAchines Layer-1}
	\centering
	\includegraphics[width=0.9\textwidth]{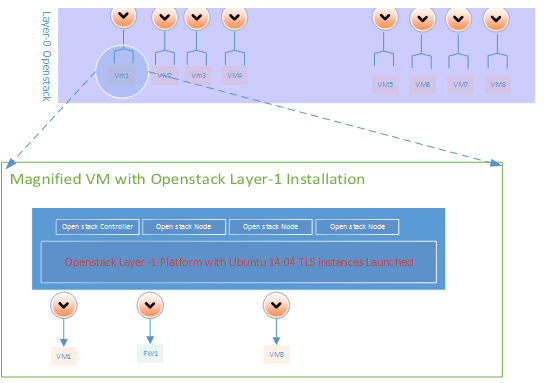}
\end{figure}
Similar to Layer 0, the Layer 1 Openstack platform set up will have all necessary components such as nova, neutron, openDaylight etc. The Remaining 12 nodes will have Openstack nova, neutron agents along with OpenDaylight agents. 
Once the Layer 1 controller and Layer 1 nodes are stacked, they can be monitored using Layer 1 dashboard, and additional VM can be launched on layer 1 Openstack platform.\\
\begin{figure}[h]
	\caption{Blades Connectivity}
	\centering
	\includegraphics[width=1.0\textwidth]{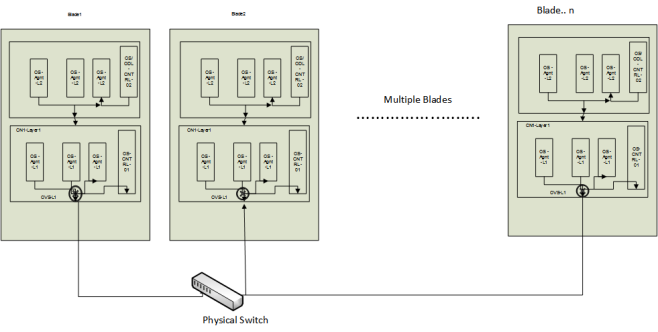}
\end{figure}

\subsection{Physical Environment}

The blades are connected together via a high speed internal physical network. Below diagram(Figure 5) shows the existing setup of the environment .Currently 4 blades are used and virtual machines are launched on these blades.The remaining Blade servers are not currently in use. The implementation is done in two layers. In the first layer, OpenStack is installed. Later on top of this layer and with the help of controller node another layer was created. This implementation of  the nodes in two layers will show the tree implementations of nodes, which is mentioned earlier in section 6.1. By integrating the nodes in two layers the examined situation will be more complicated and will burst the nodes both vertically and horizontally.

\subsection{MO-UFLP implementation}
MO-UFLP has been implemented using Java technologies and is implemented as a standalone Java program, which can be executed  via a shell script or in batch mode.NetBeans IDS 8.0.1 was used for implementing this solution in Java. We used JGAP framework, which is the Genetic Programming and Genetic Algorithms framework . It provides basic genetic mechanisms that can be easily used to apply evolutionary principles to problem solutions[5]
The cloud network is represented in a graph.

Once MO-UFLP algorithm is executed successfully and output.txt is generated, it can be fed into a specially designed Java custom scheduler to implement the topology mentioned in the output.txt file. Based on the output.txt, the program will:\\

\begin{enumerate}
	\item 	Create the network topology mentioned in the output.txt
	\item 	Automatically create and boot the VMs mentioned in the output.txt (including security Machines)
	\item 	Connect the newly created VM in to the network
	\item 	Create OpenDaylight flow rules to re-direct traffic
\end{enumerate}
This program assumes that there are no virtual machines created in the environment. Therefore, it is important to run this program in a clean environment.
\begin{figure}[h]
	\caption{OpenStack Layers and Scheduler}
	\centering
	\includegraphics[width=0.9\textwidth]{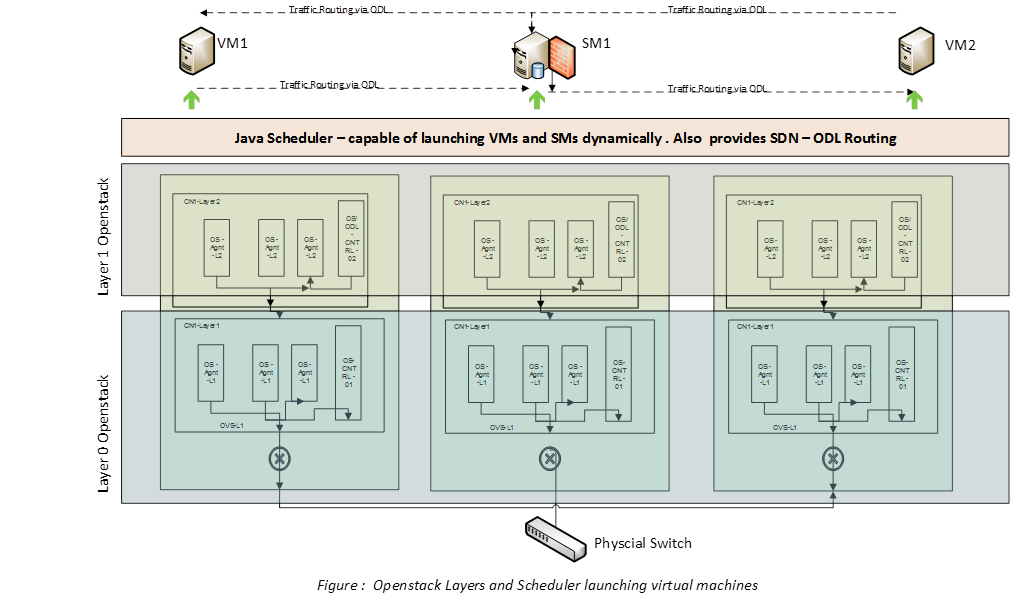}
\end{figure}

The graph will be given to the program as an input file. The implementation will read the number of total nodes, the number of security machines types required as well as the number of evolutions and the number of sources and destinations. Then the input file will include all the edges with their weight. Once the whole file is read, the graph is saved as a list of nodes and edges. The output will give the locations of security machines deployed for the current chromosome with the global cost.

\section{Conclusion}
In this paper, we presented an extension of the existing approach to the optimal placement of virtual security appliances in a pre-defined network setting. The approach proposed by Bouet [1] only considered one security appliance, we extended his approach to several virtual security appliances. We conducted several simulation tests showing good performances of our approach. To show the feasibility, we implemented our approach using SDN and virtual security appliances and integrated it into OpenStack. This extension adapts UFLP algorithm to real world situations where several middle boxes need to be deployed to satisfy security needs for the applications deployed in the cloud.

We believe our approach opens the path for further use of UFLP and genetic algorithms in more complex scenarios where many middle boxes need to be deployed in cloud infrastructure.

\section{Acknowledgments}
This research was supported and supervised by Dr.Makan Pourzandi [makan.pourzandi@concordia.ca] and Dr.Yosr Jarraya [y\_jarray@encs.concordia.ca] from Concordia University whom provided their insight and expertise for this project.

\newpage
\bibliography{latex-sample}
\item $[1]$ Bouet, Mathieu, Jeremie Leguay, and Vania Conan. "Cost-based placement of virtualized Deep Packet Inspection
functions in SDN." Military Communications Conference, MILCOM 2013-2013 IEEE. IEEE, 2013.
\item $[2]$ \url{http://www.openstack.org/}
\item $[3]$ \url{https://opensource.com/resources/what-is-openstack}
\item $[4]$ \url{http://docwiki.cisco.com/wiki/CSR1kv_in_Openstack/Neutron_Kilo_Release}
\item $[5]$ K. Meffert et al., “JGAP - Java Genetic Algorithms and Genetic Programming Package.” [Online]. Available: \url{http://jgap.sourceforge.net/}
\bibliographystyle{unsrt}

\end{document}